\documentclass[aps,showkeys,twocolumn,showpacs,superscriptaddress]{revtex4}
\usepackage{amsmath,amssymb}
\usepackage{graphics,graphicx}
\usepackage{dcolumn,bm}
\newcommand{\ictp}
{\affiliation{CMSP Section,
The Abdus Salam International Centre for Theoretical Physics,
Strada Costiera 11, Trieste I-34014, Italy.}}
\newcommand{\cu}
{\affiliation{Department of Physics, University of Calcutta, 
92 Acharya Prafulla Chandra Road, Kolkata 700009, India.}}

\begin{document}

\title{Agent dynamics in kinetic models of wealth exchange}

\author{Arnab Chatterjee}%
\email[Email: ]{achatter@ictp.it}
\ictp
\author{Parongama Sen}%
\email[Email: ]{psphy@caluniv.ac.in}
\cu

\begin{abstract}
We study the dynamics of individual agents in some kinetic models of wealth exchange,
particularly, the models with savings. For the model with uniform savings,
agents perform simple random walks in the ``wealth space''. On the other hand,
we observe ballistic diffusion in the model with distributed savings. There is 
an associated skewness in the gain-loss distribution which explains the
steady state behavior in such models.
We find that in general an agent gains while interacting with an agent
with a larger saving propensity. 

\end{abstract}


\maketitle
\section{Introduction}\label{sec:1}
The distribution of wealth among individuals in an economy has been a
very important area of research in economics, for more than a 
century~\cite{Pareto:1897,Mandelbrot:1960,EWD05,ESTP,SCCC,Yakovenko:RMP}. 
The same holds for income distribution in any society.
Detailed analysis of the income distribution~\cite{EWD05,ESTP,SCCC} so far
indicate that for large income $m$,
\begin{equation}
\label{par}
P(m) \sim m^{-(1+\nu)},  
\end{equation}
where $P$ denotes the number density of people with income 
or wealth $m$. 
The power law in income and wealth distribution is
named after Pareto and the exponent $\nu$ is called the Pareto exponent.
The tail of the income distribution indeed follows the above
mentioned behavior and the value of the Pareto exponent $\nu$ is generally
seen to vary between 1 and 
3~\cite{EWD05,SCCC,Yakovenko:RMP,datapap}.
For any country, it is well
known that typically less than $10 \%$ of the population 
possesses about $40 \%$ of the total wealth and they follow
the above law, while the rest of the low income population, 
follow a different 
distribution~\cite{EWD05,SCCC,Yakovenko:RMP,datapap,marjitIspolatov,Dragulescu:2000}.

According to physicists, the
regular patterns observed in the income (and wealth) distribution 
are indicative of a natural law for the statistical properties of a
many-body dynamical system representing the entire set of economic 
interactions in a society, analogous to those previously derived for
gases and liquids. By viewing the economy as a thermodynamic system,
one can identify the income distribution with the distribution of
energy among the particles in a gas. This has 
led to several new attempts at explaining them, particularly, 
a class of kinetic exchange models have provided a simple
mechanism for understanding the unequal accumulation of assets.
These models are simple from the perspective of economics and
implement the key factors in socioeconomic
interactions that results in very different societies converging to
similar forms of unequal distribution of resources (see 
Refs.~\cite{EWD05,ESTP}, for a collection of 
large number of technical papers in this field).

In this paper, we consider the dynamics of individual agents.
We discuss the money distribution of agents, given a particular
value of saving propensity. In models with distributed savings, we look
at  a tagged agent, and compute the distribution of money
gained or lost in each interaction. We project the gain-loss 
behavior into a walk in one dimension in a so called ``wealth space''. Our numerical
simulations suggest evidence of ballistic diffusion.

\section{Gas-like models}
\label{sec:idealgas}
In analogy to two-particle collision process which results in a change 
in their individual kinetic energy or momenta, income exchange 
models may be defined using two-agent interactions:
two randomly selected agents exchange 
money by some pre-defined mechanism. The exchange process 
does not depend on previous exchanges, hence it is a 
Markov process:
\begin{equation}
\left( \begin{array}{c}
m_i (t+1) \\
m_j (t+1)
\end{array}
\right) = {\mathcal M}
\left( \begin{array}{c}
m_i (t) \\
m_j (t)
\end{array}
\right)
\label{matrixM}
\end{equation}
where $m_i (t)$ is the income of
agent $i$ at time $t$ and the 
collision matrix ${\mathcal M}$ defines the exchange mechanism.

In this class of models, one considers a system
with $N$ agents (individuals or corporates) and total money $M$.
This is a closed economic system where $N$ and $M$ are fixed 
(microcanonical ensemble), which corresponds to no migration
or production in the system
where the only economic activity is confined to trading.
Another way of looking at this is to consider slow rates 
of growth or decay. Thus the microscopic time scale (of trading)
is much smaller in comparison to the time scale at which
the economy experiences growth or collapse.

In any trading, a pair of traders $i$ and $j$ exchange their money
\cite{marjitIspolatov,Dragulescu:2000,Chakraborti:2000,Chatterjee:rev,Chakrabarti:2010}, 
locally conserve it, while
nobody ends up with negative money ($m_i(t) \ge 0$, i.e, debt not allowed):
\begin{equation}
\label{mdelm}
m_i(t+1) = m_i(t) + \Delta m; \  m_j(t+1) = m_j(t) - \Delta m.
\end{equation}
Time ($t$) changes by one unit after each trading.

The simplest case (DY model hereafter) considers a random fraction of total money
to be shared~\cite{Dragulescu:2000}.
The steady-state ($t \rightarrow \infty$) money follows a Gibbs distribution:
$P(m)=(1/T)\exp(-m/T)$; $T=M/N$.
This result is robust and 
is independent of the topology of the (undirected)
exchange space, be it regular lattice, fractal or 
small-world~\cite{Chatterjee:rev}.

Savings is an important ingredient in a trading process~\cite{Samuelson:1980}.
A saving propensity factor $\lambda$ was introduced in the random 
exchange model~\cite{Chakraborti:2000}, where each trader
at time $t$ saves a fraction $\lambda$ of its money $m_i(t)$ and trades
randomly with the rest:
\begin{equation}
\label{fmi}
m_i(t+1)=\lambda m_i(t) + \epsilon_{ij} \left[(1-\lambda)(m_i(t) + m_j(t))\right],
\end{equation}
\begin{equation}
\label{fmj}
m_j(t+1)=\lambda m_j(t) + (1-\epsilon_{ij}) \left[(1-\lambda)(m_i(t) + m_j(t))\right],
\end{equation}
$\epsilon_{ij}$ being a random fraction, coming from the stochastic nature
of the trading.

In this model (CC model hereafter), 
the steady state distribution $P(m)$ of money is 
decaying on both sides with the most-probable money per agent shifting away
from $m=0$ (for $\lambda =0$) to $M/N$ as 
$\lambda \to 1$ ~\cite{Chakraborti:2000}. 
This model has been argued to resemble a Gamma 
distribution~\cite{Patriarca:2004,Repetowicz:2005,Lallouache:2010},
while the exact form of the distribution
for this model is still unknown.
A very similar model was proposed by 
Angle~\cite{Angle} several years back in sociology journals,
the numerical simulations of which fit well to Gamma distributions.


In a real society or economy, 
saving $\lambda$ is very inhomogeneous.
The evolution of money in a corresponding model (CCM model hereafter) 
can be written as~\cite{Chatterjee:2004}:
\begin{equation}
\label{mi}
m_i(t+1)=\lambda_i m_i(t) + \epsilon_{ij} \left[(1-\lambda_i)m_i(t) + (1-\lambda_j)m_j(t)\right],
\end{equation}
\begin{equation}
\label{mj}
m_j(t+1)=\lambda_j m_j(t) + (1-\epsilon_{ij}) \left[(1-\lambda_i)m_i(t) + (1-\lambda_j)m_j(t)\right].
\end{equation}
It looks similar to the CC model, except that
$\lambda_{i}$ and $\lambda_{j}$, the saving propensities of agents 
$i$ and $j$, are different. The agents have saving
propensities, distributed randomly and independently as $\Lambda(\lambda)$,
such that $\Lambda(\lambda)$ is non-vanishing as $\lambda \to 1$;
$\lambda_i$ is quenched
for each agent ($\lambda_i$ are independent of trading or $t$).
The actual asset distribution $P(m)$ in such a model depends on the 
form of $\Lambda(\lambda)$, but for all of them the asymptotic form of 
the distribution will become Pareto-like (Eqn.~(\ref{par})).
For uniform distribution, $\Lambda(\lambda) = 1$,  $\nu=1$. 
However, for distributions
\begin{equation}
 \Lambda(\lambda) \propto (1-\lambda)^{\delta}, 
\end{equation}
$P(m) \sim m^{-(2+\delta)}$~\cite{Chatterjee:2004,Mohanty:2006,Chatterjee:rev}.
In the CCM model, agents with higher saving propensity tend to hold
higher average wealth, which is justified by the fact that the saving
propensity in the rich population is always high~\cite{Dynan:2004}.
Analytical understanding of CCM model has been possible until now
under certain approximations~\cite{Chatterjee:2005}, and mean-field 
theory~\cite{Mohanty:2006,Kargupta}  suggests that the agent with
saving $\lambda$ possesses wealth $m(\lambda)$ in the steady state,
where
\begin{equation}
m(\lambda) = C/(1-\lambda),
\label{av-wealth}
\end{equation}
with $C \propto 1/\log(N)$.
The precise analytical formulation of the above models have been
recently considered with success, and the results have been
derived in most cases~\cite{Toscani}.
There have been recent efforts in analyzing the CCM model
at a microscopic scale~\cite{Chatterjee:2010}.
Variations of these models have been considered in several
manners to obtain similar (e.g., for annealed case~\cite{ecoanneal}) or different (as on networks) results 
\cite{Chatterjee:2009,Chakraborty:2010}.
It has also been shown that the results are invariant  even if one considers
a cononical ensemble~\cite{Basu:2008}.
Some microeconomic formulations have also proved to be useful~\cite{ChakrabartiASBK}.

\section{Studying dynamics of agents}
\label{sec:dynamics}

In the DY and CC models, agents are homogeneous. DY model is nothing but a special
case of the CC model where $\lambda=0$. 
In these models, looking at individual agents and the whole system are equivalent. 
On the contrary, the presence of the distributed saving propensity 
(quenched disorder) in CCM model gives it a rich structure and 
calls for a careful look at the local scale (at the level of individuals) 
besides computing global quantities.

\begin{figure}
\centering \includegraphics[width=8.7cm]{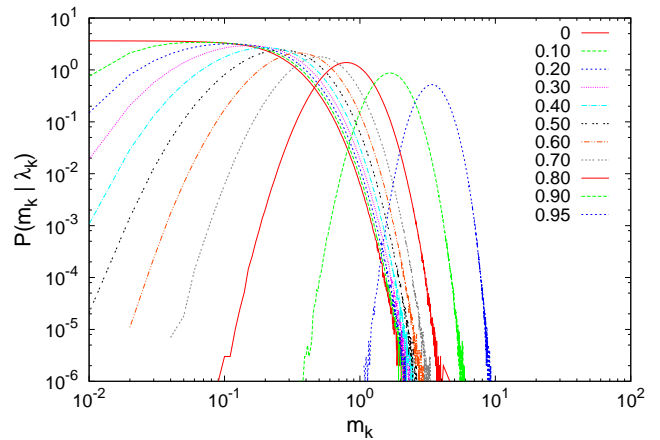}
\caption{
Distribution $P(m_k | \lambda_k)$ of money $m_k$ for the tagged agent $k$ with
a particular value of savings $\lambda_k$ in the CCM model with uniformly
distributed savings ($\delta=0$). 
The data is shown for a system of $N=256$ agents.
}
\label{fig:moneytag}
\end{figure}
In this study, we perform extensive numerical simulations with system of $N$ agents,
with uniform distribution $\Lambda(\lambda)=1$, bounded above by $1-1/N$.
We look at the dynamics of a tagged agent $k$, having a saving propensity $\lambda_k$,
in a pool of $N$ agents distributed according to a quenched $\Lambda(\lambda)$.
We try to see how the individual distributions $P(m_k | \lambda_k)$ look like 
(see Fig.~\ref{fig:moneytag}). As reported 
elsewhere~\cite{Chatterjee:2004,Chatterjee:rev,papertag1,papertag2},
the agents with smaller values of savings $\lambda_k$ barely have money of the order of average 
money in the market. On the other hand, agents with high saving propensity $\lambda_k$
possess money comparable to the average money in the market, and in fact, for the richest
agent, the distribution extends almost upto the total money $M$.

\subsection{Distribution of change in wealth}

\begin{figure}
\centering \includegraphics[width=8.7cm]{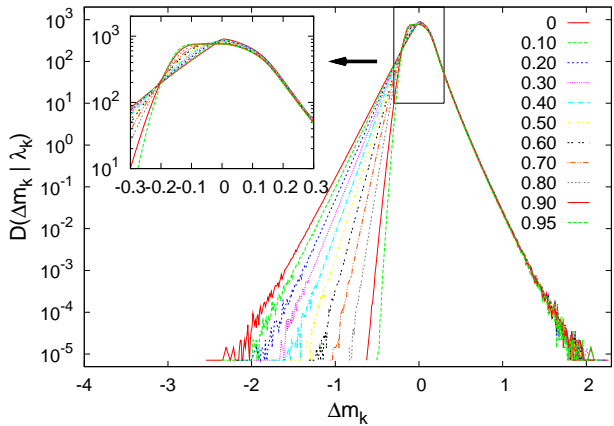}
\caption{
Distribution $D(\Delta m_k | \lambda_k)$ of money difference $\Delta m_k$
for the tagged agent $k$ with a particular value of savings $\lambda_k$ 
in the CCM model with uniformly distributed savings ($\delta=0$). 
The data is shown for a system of $N=256$ agents. 
The inset shows that for higher $\lambda$, probability of
losses become larger in the region $-0.2 < \Delta m_k < 0$.
}
\label{fig:moneydiff}
\end{figure}

Upon trading with another agent $l$, the money of the tagged agent $k$
changes by an amount
\begin{equation*}
 \Delta m_k = m_k(t+1)-m_k(t)=-\left(m_l(t+1)-m_l(t)\right).
\end{equation*}
We compute the distributions $D(\Delta m_k | \lambda_k)$ in the steady state, 
given that agent $k$ has a saving propensity $\lambda_k$ (Fig.~\ref{fig:moneydiff}).
We observe this distribution to have asymmetries both for small and large values
of saving propensities $\lambda_k$. 

Total money remains constant in the steady state for any agent.
An agent with a relatively higher $\lambda$ incurs losses which 
are considerably small in magnitude. This immediately suggests
that agents with larger savings must be having more exchanges where  
losses, however small, occur. The magnified portion of the
distribution shows that it is really so (shown in the inset of Fig. \ref{fig:moneydiff}).

\subsection{Walk in the wealth space: Definition}
\label{ssec:walk}

To investigate the dynamics at the microscopic level, 
one can conceive of  a walk for the agents in 
the so called  ``wealth space'', in which each agent 
walks a step forward when she gains and one step backwards if she 
incurs a loss.
The walks are correlated in the sense that when two agents interact, if
one takes a step forward, the other has to move backward. On the other
hand, two agents can interact irrespective of their positions in the
wealth space unlike Brownian particles.

Once the system is in the steady state, we define $x(t+1)=x(t)+1$
if our tagged agent gains money, and $x(t+1)=x(t)-1$ if she loses.
In other words, $x(t)$ performs a walk in one dimension.
Without loss of generality we start from origin ($x(0)=0$),
and we insist $t=0$ is well within the steady state.
We investigate the properties of this walk by computing the mean displacement
$\langle x(t) \rangle$, and the mean square displacement 
$\langle x^2(t) \rangle - \langle x(t) \rangle^2$.

Actually, one can also consider a walk for a tagged agent
where the increments (i.e., step lengths) are
the money gained or lost at each step, but the exponential distribution
obtained for such step lengths (Fig.~\ref{fig:moneydiff}) indicates that it will
be  simple diffusion like~\cite{Majumdar}.

\subsection{Results for the walk}
\label{ssec:walkresults}

\begin{figure}
\centering \includegraphics[width=8.7cm]{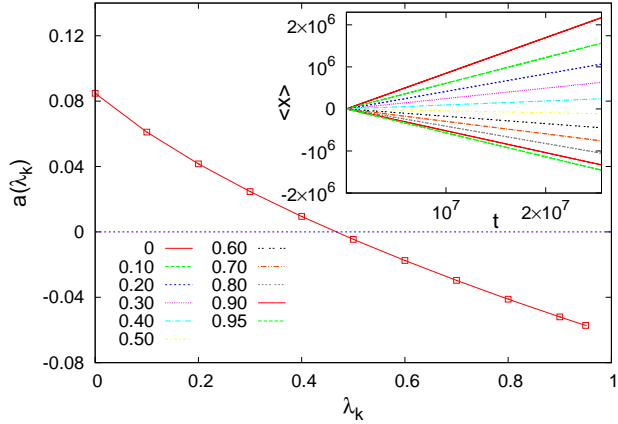}
\caption{
Measures for the gain-loss walk: 
the inset shows $\langle x \rangle$ with time
for different values of savings $\lambda_k$, showing
the drifts. The slopes $a(\lambda_k)$ are also shown.
The estimate of $\lambda_k^*$ is approximately $0.469$.
The data is shown for a system of $N=256$ agents.
}
\label{fig:scwalk2}
\end{figure}

\begin{figure}
\centering \includegraphics[width=8.7cm]{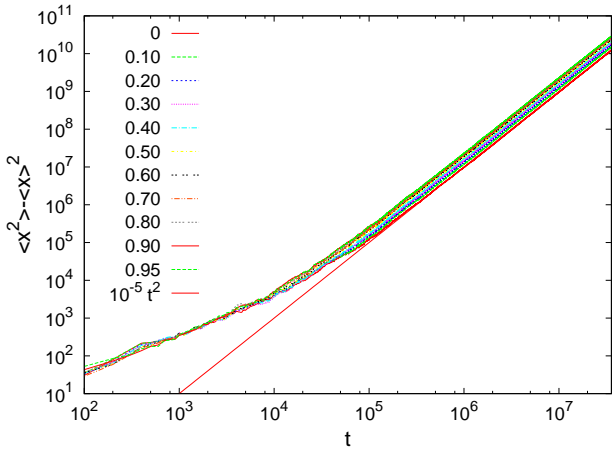}
\caption{
Measures for the gain-loss walk: time variation of 
$\langle x^2 \rangle - \langle x \rangle^2$ for different values 
of savings $\lambda_k$, and a guide to $t^2$.
The data is shown for a system of $N=256$ agents.
}
\label{fig:scwalk1}
\end{figure}

For the CC model, for any value of the fixed saving propensity $\lambda$, we
obtain a conventional random walk in the sense  $\langle x(t) \rangle $ is zero and 
 $\langle x^2(t) \rangle - \langle x(t) \rangle^2 \sim t$.
However, for the CCM model, results are quite different.
It is found that $\langle x(t) \rangle$ has a drift,
$\langle x(t) \rangle \sim a(\lambda_k) t$.  $a(\lambda_k)$
varies continuously with $\lambda_k$, taking positive to negative values
as one goes from low to high values of savings $\lambda_k$ (see inset of Fig.~\ref{fig:scwalk2}) respectively.
It is obvious that for some $\lambda_k^*$, there is no drift, $a(\lambda_k^*)=0$.
$\lambda_k^*$ is estimated to be about $0.469$ by interpolation method.
On the other hand $\langle x^2 \rangle - \langle x \rangle^2 \sim t^2$
for all $\lambda_k$, which is a case of ballistic diffusion (Fig.~\ref{fig:scwalk1}).

\begin{figure}
\centering \includegraphics[width=9.0cm]{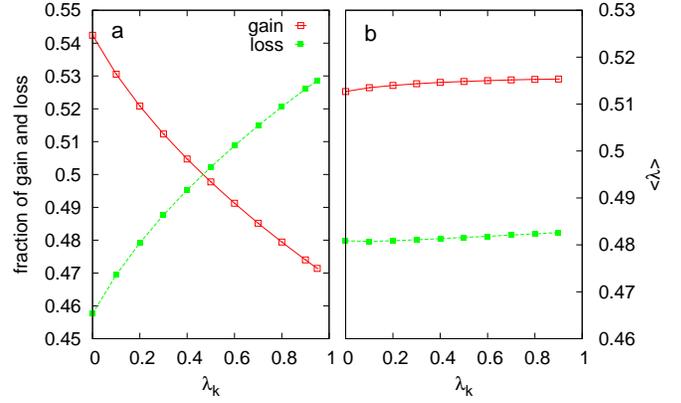}
\caption{
(a) Plot of probabilities of gain and loss for different values 
of savings propensity $\lambda_k$ of the tagged agent. 
The data is shown for a system of $N=256$ agents.
(b) The plot of the average value $\langle \lambda \rangle$ when 
a gain/loss is being incurred  shown against $\lambda_k$ of the tagged agent.
}
\label{fig:gainloss}
\end{figure}

The negative or positive drifts of the walks indicate that the 
probabilities of gain and loss are not equal for any agent in general.
Plotting the fraction of times the tagged agent gains/loses in
Fig.~\ref{fig:gainloss}(a), it is indeed found that an agent with a 
smaller $\lambda$ gains with more probability while the opposite
happens for agents with larger $\lambda$. Indeed, the intersection of the
two curves is the point $\lambda_k^*$ where the probabilities are equal and 
the corresponding walk should show $\langle x \rangle =0$.  It 
is difficult, however, to detect numerically $\lambda_k^*$ exactly, 
which lies close to $0.47$, and check whether 
an agent with  $\lambda_k^*$  behaves like a conventional random walker
or shows ballistic diffusion.
Simulations using  values of $\lambda$ even very   
  close to $0.47$ always show ballistic behavior.

In order to explain the above results, we investigate at 
a finer level the walk when the tagged agent with $\lambda_{k}$
interacts with another agent with saving $\lambda$. First, we calculate the
average $\langle \lambda \rangle$ when the tagged agents loses or gains and find that
for a gain, one has to interact with an agent with a higher $\lambda$ in
general. This is shown in Fig.~\ref{fig:gainloss}(b). In fact, the
average value is very weakly dependent on $\lambda_k$ and significantly 
greater/less than 0.5 for a gain/loss.
This is contrary to the expectation that gain/loss does not depend on the
saving propensities of the interacting agents.

\begin{figure}
\centering \includegraphics[width=8.7cm]{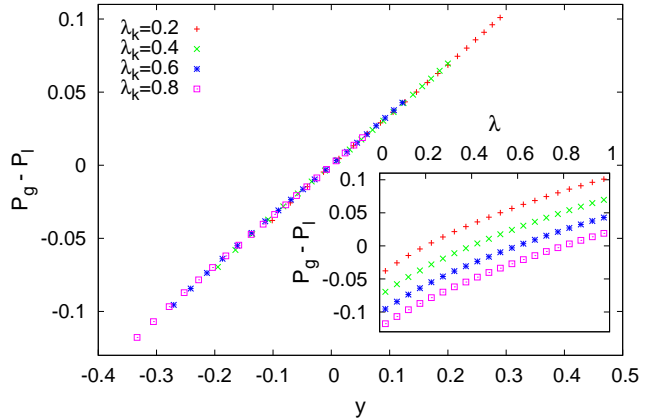}
\caption{Data collapse for $\lambda_k = 0.2, 0.4, 0.6 $ and $0.8$ is shown for 
$P_g-P_l$ versus the scaled variable $y=\frac{\lambda -\lambda_{k}}{1.5+\lambda_{k} + \lambda}$ for $N = 256$.
The inset shows the unscaled data.
}
\label{fig:prob-gainloss}
\end{figure}

Having obtained evidence that gain/loss depends on the 
  interacting agents' saving propensities,
we  compute the probability of gain and loss, $P_g$ and $P_l$ respectively,
as a function of $\lambda$ for the agent with saving $\lambda_k$.
 The data shows that   indeed 
an agent gains with higher probability while interacting with an agent with  
$\lambda > \lambda_k$ and vice versa. In fact, the data  for different $\lambda_{k}$
collapse when  $P_g - P_l$  are 
 plotted against a scaled variable  
$y = \frac{\lambda -\lambda_{k}}{1.5+\lambda_{k} + \lambda}$ 
as shown in Fig.~\ref{fig:prob-gainloss}
indicating  a linear variation  with $y$, i.e.,
\begin{equation}
\label{eq:pgpl}
P_g - P_l = {\rm const} \frac{\lambda -\lambda_k}{1.5+\lambda_k + \lambda}.
\end{equation}
We have checked that there is hardly any finite size effect on the collapse 
in the sense that the similarly scaled data for $N=100$ collapse exactly on those for $N=256$.
 An agent with a high value of $\lambda$ will interact with a higher probability with agents whose  saving propensities are  lesser, causing a loss of money. Therefore
in the wealth space it will have tendency to take more steps in the 
negative direction. This explains the negative drift for 
large $\lambda$. 

It is possible to estimate the value of $\lambda_k^*$ using Eq.~(\ref{eq:pgpl})
utilizing   the fact  that the integrated value of $P_g -P_l$ over all 
$\lambda$ should be zero for $\lambda_k = \lambda_k^*$. This gives
\begin{equation}
\label{eq:pgpl-int}
1-(1.5 +2\lambda_k^*)\left[\log(2.5 +\lambda_k^*)-\log(1.5 +\lambda_k^*)\right] =0,
\end{equation}
solving which we get $\lambda_k^* \simeq 0.4658$ which is consistent with
the earlier observations.

It may be added here that in principle the probability of gain or of loss while two agents interact,
can be calculated from the money distribution. In the CCM model,
when two agents with  money $m_1$ and $m_2$ and saving propensities 
$\lambda_1$ and $\lambda_2$ respectively, interact, the difference in money 
before and after transcation for, say, the second agent is given by 
$\left[ (1-\lambda_1)m_1 - (1-\lambda_2)m_2\right] /2$.
Therefore for the second agent to lose, $m_2$ must be greater than 
$m^\prime = \frac{m_1(1-\lambda_1)}{(1-\lambda_2)}$
and the corresponding 
probability  is given by
\begin{equation}
\int_0^{M} P(m_1 | \lambda_1)dm_1 \int^M_{m^\prime} P(m_2 | \lambda_2) dm_2.
\label{integral-eq}
\end{equation}
However, the exact form of the money distribution is not 
known~\cite{Basu:2008} for the CCM case. For the CC model,
$\lambda = \lambda_1 = \lambda_2$ and letting $M \to \infty$, the above integral becomes 
\begin{eqnarray}
\int_0^{\infty} P(m_1 | \lambda)dm_1 \int^\infty_{m_1} P(m_2 | \lambda) dm_2 \nonumber \\
=\int_0^{\infty} P(m_1 | \lambda)[1-\tilde P(m_1 | \lambda)]dm_1, 
\label{integral-eq-cc}
\end{eqnarray}
where $\tilde P(m) = \int_0^m P(m)dm$ is the cumulative distribution of money.
Since $P =\frac{ \partial {\tilde P}}{\partial m}$,
the R.H.S of Eqn.~(\ref{integral-eq-cc}) is equal to 1/2 independent of the form of $P(m | \lambda)$.
Thus in the CC case we find equal probability of gain or loss leading to 
a simple random walk. 
In the CCM, however, the results are expected to be dependent on 
$\lambda_1, \lambda_2$ as  (\ref{integral-eq}) indicates.



\section{Discussions}
\label{sec:disc}
The analogy with a gas like many-body system has led
to the formulation of the kinetic exchange models of markets.
The random scattering-like dynamics of money (and wealth) in a closed 
trading market, in analogy with energy conserved exchange models,
reveals interesting features.
Self-organization is a key emerging feature of these simple
models when saving factors are introduced.
These models produce asset distributions resembling those observed in reality,
and are quite well studied now~\cite{Chatterjee:rev}. 
These have prospective applications in other spheres of social science, as 
in application in policy making and taxation,
and also physical sciences, possibly in designing desired energy 
spectrum for different types of chemical reactions.

In this paper, 
we  have looked  at the dynamics of agents at the scale of individuals.
We study the distribution of money for a tagged agent given a particular value
of saving propensity. We also analyze the distribution of money differences
in successive exchanges. 
We conceived of a walk in an abstract space performed
by the agents in different kinetic gas like models which reveals the 
characteristics of gains and losses made by the agents. 
Specifically, considering only whether an agent gains or loses, 
a walk can be defined in the wealth space, which is a random walk for the CC model 
while for the CCM model it is  found to be ballistic
in nature for generic values of $\lambda$.

On studying the dynamics at a microscopic level, we find that
an agent gains with a higher probability when interacting with 
another agent with a larger $\lambda$. Thus one would expect, that for $\lambda=0.5$,
there will be equal gains and losses which would give rise zero drift 
in the corresponding walk  in
the wealth space for the CCM model. The value of $\lambda_k^*$ 
for which we do get such a result is quite close to this estimate. 
An accurate estimate of $\lambda_k^* \simeq 0.4658$ is obtained
by considering  the scaling behavior of (probabilities of) gains and losses.

In the CCM model, our study leads to the
discovery of the way  gains and losses are dependent on the
saving propensities of the agents, which cannot be
arrived at  using  any  existing results, either analytical or numerical.
Using Eqn.~(\ref{av-wealth}), 
one can calculate the average money exchanged between two agents
which turns out to be independent of their saving propensities.
Hence the intriguing question that remains to be solved is why the probability 
of gain over loss  depends  on the savings of the interacting agents.

\begin{acknowledgments}
The authors thank 
B.~K.~Chakrabarti, S.~S.~Manna
for some useful comments and discussions.
AC was supported by the ComplexMarkets E.U. STREP project  
516446 under FP6-2003-NEST-PATH-1.
PS is supported by DST project no SR/S2/CMP-56/2007.
\end{acknowledgments}


\end{document}